\documentclass{ifacconf}

\usepackage{graphicx}      
\usepackage{natbib}        
\usepackage{amsmath}
\usepackage{amsfonts}
\usepackage{xcolor}

\newcommand{\R}{\mathbb{R}}
\newcommand{\C}{\mathbb{C}}
\newcommand{\N}{\mathbb{N}}
\newcommand{\PP}{\mathbb{P}}

\newcommand{\LL}{\mathbb{L}}

\DeclareMathOperator{\rank}{rank}

\newtheorem{definition}[thm]{Definition}
\newtheorem{problem}[thm]{Problem}
\newtheorem{remark}[thm]{Remark}
\usepackage{tikz}
\usepackage{tkz-graph}
\usepackage{graphicx}      

\usepackage[american]{circuitikz}
\makeatletter
\ctikzset{bipoles/vsources/minus/sign/rotation/.initial=0}
\pgfcircdeclarebipole{}{\ctikzvalof{bipoles/vsourceam/height}}{vsourceAM}{\ctikzvalof{bipoles/vsourceam/height}}{\ctikzvalof{bipoles/vsourceam/width}}{
	\pgfsetlinewidth{\pgfkeysvalueof{/tikz/circuitikz/bipoles/thickness}\pgfstartlinewidth}
	\pgfpathellipse{\pgfpointorigin}{\pgfpoint{0}{\pgf@circ@res@up}}{\pgfpoint{\pgf@circ@res@left}{0}}
	\pgfusepath{draw}
	\ifpgf@circ@oldvoltagedirection
	\pgftext[bottom,rotate=90,y=\ctikzvalof{bipoles/vsourceam/margin}\pgf@circ@res@down]{$+$}
	\pgftext[top,rotate=90,y=\ctikzvalof{bipoles/vsourceam/margin}\pgf@circ@res@up]{$-$}
	\else
	\pgftext[bottom,rotate=90,y=\ctikzvalof{bipoles/vsourceam/margin}\pgf@circ@res@down]{\rotatebox{\ctikzvalof{bipoles/vsources/minus/sign/rotation}}{$-$}}
	\pgftext[top,rotate=90,y=\ctikzvalof{bipoles/vsourceam/margin}\pgf@circ@res@up]{$+$}
	\fi
}

\pgfcircdeclarebipole{}{\ctikzvalof{bipoles/cvsourceam/height}}{cvsourceAM}{\ctikzvalof{bipoles/cvsourceam/height}}{\ctikzvalof{bipoles/cvsourceam/width}}{
	\pgfsetlinewidth{\pgfkeysvalueof{/tikz/circuitikz/bipoles/thickness}\pgfstartlinewidth}
	\pgfpathmoveto{\pgfpoint{\pgf@circ@res@left}{\pgf@circ@res@zero}}
	\pgfpathlineto{\pgfpoint{\pgf@circ@res@zero}{\pgf@circ@res@up}}
	\pgfpathlineto{\pgfpoint{\pgf@circ@res@right}{\pgf@circ@res@zero}}
	\pgfpathlineto{\pgfpoint{\pgf@circ@res@zero}{\pgf@circ@res@down}}
	\pgfpathlineto{\pgfpoint{\pgf@circ@res@left}{\pgf@circ@res@zero}}
	\pgfusepath{draw}
	
	\ifpgf@circ@oldvoltagedirection
	\pgftext[bottom,rotate=90,y=\ctikzvalof{bipoles/cvsourceam/margin}\pgf@circ@res@left]{$+$}
	\pgftext[top,rotate=90,y=\ctikzvalof{bipoles/cvsourceam/margin}\pgf@circ@res@right]{$-$}
	\else
	\pgftext[bottom,rotate=90,y=\ctikzvalof{bipoles/cvsourceam/margin}\pgf@circ@res@left]{\rotatebox{\ctikzvalof{bipoles/vsources/minus/sign/rotation}}{$-$}}
	\pgftext[top,rotate=90,y=\ctikzvalof{bipoles/cvsourceam/margin}\pgf@circ@res@right]{$+$}
	\fi
}

\begin{document}
\begin{frontmatter}

\title{Informativity for data-driven model reduction through interpolation\thanksref{footnoteinfo}
}

\thanks[footnoteinfo]{This extended abstract is based on research developed in the DSSC Doctoral Training Programme, co-funded through a Marie Skłodowska-Curie COFUND (DSSC 754315).}

\author[First,Second,Third]{Azka M. Burohman} 
\author[First,Second]{Bart Besselink} 
\author[First,Third]{Jacquelien M. A. Scherpen}
\author[First,Second]{M. Kanat Camlibel}

\address[First]{Jan C. Willems Center for Systems and Control, University of Groningen, The Netherlands.}
\address[Second]{Bernoulli Institute for Mathematics, Computer Science and Artificial Intelligence, University of Groningen, The Netherlands. (e-mail:\{a.m.burohman, b.besselink, m.k.camlibel\}@rug.nl)}
\address[Third]{Engineering and Technology Institute Groningen, University of Groningen, The Netherlands. (e-mail: j.m.a.scherpen@rug.nl)}

\begin{abstract}                
A method for data-driven interpolatory model reduction is presented in this extended abstract. This framework enables the computation of the transfer function values at given interpolation points based on time-domain input-output data only, without explicitly identifying the high-order system. Instead, by characterizing the set of all systems explaining the data, necessary and sufficient conditions are given under which all systems in this set share the same transfer function value at a given interpolation point. After following this so-called data informativity perspective, reduced-order models can be obtained by classical interpolation techniques. An example of an electrical circuit illustrates this framework.
\end{abstract}


\end{frontmatter}

\section{Introduction}
High-order dynamical systems naturally appear when modeling complex phenomena, e.g., in modern engineering systems. This results from either the inherent complexity of the systems or from the discretization of partial differential equations. The approximation of the input-output behavior of such a model by a model of lower order is known as model reduction, which has become a crucial tool in the analysis and control of complex systems. 
Interpolatory techniques form a popular class of model reduction approaches, since they are numerically stable and, therefore, applicable to large-scale models. These methods are aimed at constructing a reduced-order model whose transfer function interpolates that of the original high-order model at selected interpolation points, e.g., \cite{antoulas2010interpolatory}.

Moment matching techniques form an example of interpolatory model reduction methods and were originally developed in the field of numerical mathematics, see, e.g., \cite{grimme1997krylov,feldmann1995efficient}. By exploiting Krylov subspaces and projection, these methods achieve interpolation without explicitly computing the transfer function values. Consequently, this method is well suited for interpolating also the (high-order) derivatives (moments) of the transfer function, see also \cite{gugercin2008h_2,gallivan2004model}. A thorough exposition of this topic can be found in \cite{antoulas2005approximation}.  
This well-established projection method was revisited in \cite{astolfi2010model}, providing a time-domain perspective on moment matching.

The majority of these model reduction methods rely on the availability of a state-space model or transfer function of the system to be reduced. In this extended abstract, however, we develop a \emph{data-driven} interpolation method, i.e., based exclusively on input-output measurements on the high-order system. This is motivated by the observation that, in many cases, an explicit system model (or access to it) is not available.

Data-driven model reduction has been considered before in several works. In \cite{scarciotti2017data}, building on the framework of \cite{astolfi2010model}, time-domain snapshots of input-output data are used to derive a reduced-order model by estimating its moments. In \cite{peherstorfer2016dynamic}, transfer function values at given interpolation points are estimated by exploiting the (discrete) Fourier transform. After computing the estimation of the transfer function values, the reduced-order model is computed using the Loewner method. In addition, if the state trajectories are available, then dynamic mode decomposition provides a way to find a linear model that best fits the given trajectories in the $\LL_2$ sense as in \cite{proctor2016dynamic}. And recently in \cite{monshizadeh2020amidst}, if the state trajectories are available, a lower-order model expressed in terms of data is uncovered such that its complexity matches that of the given data. Earlier results on data-driven model reduction are provided in \cite{rapisarda2011identification}, using fundamental lemma of \cite{willems2005note}. Assuming persistent excitation on the input data, a balanced realization can be extracted directly from data, and then balanced truncation can be applied to obtain a reduced-order model.

\textcolor{black}{
The contributions of this extended abstract are as follows. First, we introduce a new framework for computing transfer function values at given interpolation points without identifying the model. This perspective builds on the concept of data informativity introduced in \cite{van2019data}, which aims to find conditions on input-output data such that specific system properties can be concluded from this data. Consequently, this framework does not require persistently exciting data, which is in fact a sufficient condition for fully identifying the system (rather than only specific system properties), see \cite{verhaegen2007filtering,Ljung1999System}. Second, we provide necessary and sufficient conditions for data informativity for the computation of transfer function values at given interpolation points. Finally, we can obtain the reduced-order transfer function (and hence the reduced-order system) by application of classical interpolation methods. These methods include the use of the Loewner matrix as the main tool to achieve, e.g., a stable rational interpolation, \cite{antoulas1989problem}, and a generalized realization \cite{mayo2007framework}. Minimal rational interpolation is also considered in \cite{antoulas1990minimal} using Prony's method.}

This extended abstract is organized as follows. In Section \ref{Section_problem} we introduce the problem at a conceptual level. Subsequently, in Section \ref{Section_result} we provide the informativity condition for interpolation. Next, Section \ref{Section_example} contains an illustrative example, and in Section \ref{Section_conclusion} we give our conclusions and future works. 

\section{Problem Formulation}\label{Section_problem}
In this section we will introduce the notion of data informativity for interpolation.

Consider the discrete-time input-output system of the form
\begin{equation}\label{YU}
\begin{split}
y_{t+n} &+ \bar{p}_{n-1}y_{t+n-1} + \cdots + \bar{p}_1 y _{t+1} + \bar{p}_0y_t \\ 
&= \bar{q}_nu_{t+n} + \bar{q}_{n-1}u_{t+n-1} +  \cdots + \bar{q}_1u_{t+1} + \bar{q}_0u_t,
\end{split}
\end{equation}
where $u$ denotes the input, $y$ the output and $t \in \N$ the discrete-time variable. The parameters of (\ref{YU}) are collected in the row vector $\begin{bmatrix}
\bar{q} & -\bar{p}
\end{bmatrix}$ where
\begin{equation*}
\bar{p}=\begin{bmatrix}
\bar{p}_0 & \bar{p}_1 & \cdots & \bar{p}_{n-1}
\end{bmatrix} \in \R^{1 \times n}
\end{equation*}
and
\begin{equation*}
\bar{q}=\begin{bmatrix}
\bar{q}_0 & \bar{q}_1 & \cdots & \bar{q}_n
\end{bmatrix} \in \R^{1 \times (n+1)}.
\end{equation*}
We will refer to (\ref{YU}) as the `true' system. However, in the remainder of this paper, we assume that its parameters $\begin{bmatrix}
\bar{q} & -\bar{p}
\end{bmatrix}$ are unknown. Instead, we have access to a set of input-output data
$(U,Y)$ generated by (\ref{YU}), where
\begin{equation*}
U= \begin{bmatrix}
\bar{u}_0 & \bar{u}_1 & \cdots & \bar{u}_T
\end{bmatrix} \quad \text{and} \quad Y= \begin{bmatrix}
\bar{y}_0 & \bar{y}_1 & \cdots & \bar{y}_T
\end{bmatrix},
\end{equation*}
for some integer $T >0 $.

As the `true' system (\ref{YU}) is assumed to be unknown, we would like to characterize the set of all systems that could have generated the data $(U,Y)$. To this end, assume that $n$ is given and let
\begin{equation}\label{sys_data}
\begin{split}
y_{t+n} &+ {p}_{n-1}y_{t+n-1} + \cdots + {p}_1 y _{t+1} + {p}_0y_t \\ 
&= {q}_nu_{t+n} + {q}_{n-1}u_{t+n-1} +  \cdots + {q}_1u_{t+1} + {q}_0u_t,
\end{split}
\end{equation}
be one of such system. After defining
\begin{equation*}
{p}=\begin{bmatrix}
p_0 & p_1 & \cdots & p_{n-1}
\end{bmatrix} \in \R^{1 \times n}
\end{equation*}
and
\begin{equation*}
{q}=\begin{bmatrix}
q_0 & q_1 & \cdots & q_n
\end{bmatrix} \in \R^{1 \times (n+1)},
\end{equation*}
it is clear that (\ref{sys_data}) explains the data $(U,Y)$ (i.e., could have generated the data) if and only if
\begin{equation}\label{linear1}
\begin{bmatrix}
q & -p
\end{bmatrix} \begin{bmatrix}
H_{n}(U) \\ \bar{H}_{n}(Y)
\end{bmatrix} = \begin{bmatrix}
\bar{y}_n & \bar{y}_{n+1} & \cdots & \bar{y}_{T}
\end{bmatrix},
\end{equation}
where 
\begin{equation*}
H_{\ell}(U) = \begin{bmatrix}
\bar{u}_0 & \bar{u}_1 & \cdots & \bar{u}_{T-\ell} \\
\bar{u}_1 & \bar{u}_2 & \cdots & \bar{u}_{T-\ell+1} \\
\vdots & \vdots & \ddots & \vdots \\
\bar{u}_{\ell} & \bar{u}_{\ell+1} & \cdots & \bar{u}_{T}
\end{bmatrix} \in \R^{(\ell+1)\times (T-\ell+1)}.
\end{equation*}
is the Hankel matrix of $U$ of depth $\ell$. The Hankel matrix of the output data $Y$ is defined similarly and $\bar{H}_{n}(Y)$ equals $H_{n}(Y)$ with its last row removed. 
Now, we can define
\begin{equation}\label{Set_UY}
\Sigma_{U,Y}:=\left\lbrace  \left. \begin{bmatrix}
q & -p
\end{bmatrix} \in \R^{1 \times (2n+1)}\right| \quad (\ref{linear1}) \quad  \text{holds} \right\rbrace
\end{equation}
as the set containing all (parameterization of) systems explaining the data $(U,Y)$.
Since (\ref{YU}) generated the data, it is clear that $\begin{bmatrix}
\bar{q} & -\bar{p}
\end{bmatrix} \in \Sigma_{U,Y}$. 

\subsection{Transfer Function Values of a Discrete-Time Linear System}
Before defining the transfer function value of system (\ref{sys_data}), we introduce the forward shift operator $z$, defined as $zf_t = f_{t+1}$. This allows (\ref{sys_data}) to be written in the operational form
\begin{equation}\label{Eqn_polyyu}
\begin{split}
&\left(z^n+p_{n-1}z^{n-1}+\cdots+p_1z+p_0\right)y_t \\ & \quad \quad \quad \quad = \left(q_nz^n+q_{n-1}z^{n-1}+\cdots+q_1z+q_0\right)u_t.
\end{split}
\end{equation}

We denote the polynomial in the left and right-hand side of (\ref{Eqn_polyyu}) as
\begin{equation}\label{nota_PQ}
\begin{split}
&P(z):=z^n+p_{n-1}z^{n-1}+\cdots+p_1z+p_0, \\
&Q(z):=q_nz^n+q_{n-1}z^{n-1}+\cdots+q_1z+q_0.
\end{split}
\end{equation}

Now we are in the position to define the transfer function value of (\ref{sys_data}). 

\begin{definition}[transfer function value]\label{Def_0moment}
	Given an interpolation point $\sigma \in \C$, the number $M \in \C$ is said to be the transfer function value at $\sigma$ of the discrete-time system (\ref{sys_data}) if it is a solution to 
	\begin{equation}\label{defmoment}
	P(\sigma)M = Q(\sigma).
	\end{equation}
	In this case, we also write $M=M(\sigma)$.
	\end{definition}
\begin{remark}
	The notation in Definition \ref{Def_0moment} is a slight generalization of the  classical definition of transfer function value as it allows to define a transfer function value in case both $P(\sigma)=0$ and $Q(\sigma)=0$.
	\end{remark}

Given (\ref{nota_PQ}), condition (\ref{defmoment}) can be written as the linear equation 
\begin{equation}\label{linear2}
\begin{bmatrix}
q & -p
\end{bmatrix}\begin{bmatrix}
\gamma_n(\sigma) \\ M \gamma_{n-1}(\sigma)
\end{bmatrix} = M\sigma^n,
\end{equation}
where
\begin{equation*}
\gamma_{\ell}(z) := \begin{bmatrix}
1 \\ z \\ \vdots \\ z^{\ell}
\end{bmatrix}.
\end{equation*}
Then, the expression (\ref{linear2}) allows for defining 
\begin{equation}\label{Set_sigmaM}
\Sigma_{\sigma,M}:=\left\{ \left.\begin{bmatrix}
q & -p
\end{bmatrix}\in \R^{1 \times (2n+1)} \right| \quad (\ref{linear2}) \quad  \text{holds} \right\}
\end{equation}
as the set of all (parameterization of) systems of order $n$ that have transfer function value $M$ at $\sigma$.

\subsection{Informativity Problem for Interpolation}
Recall that we are interested in the transfer function value of the true system (\ref{YU}), but only have the input-output data $(U,Y)$ available. Given an interpolation point $\sigma \in \C$, we are interested in finding (necessary and sufficient) conditions for the data $(U,Y)$ to be sufficiently rich to allow for computing the transfer function value at $\sigma$. 

Recall that all systems with transfer function value $M$ at $\sigma$ are given by the set $\Sigma_{\sigma,M}$. However, as $\begin{bmatrix}
\bar{q} & -\bar{p}
\end{bmatrix} \in \Sigma_{U,Y}$, it is sufficient to ask for $\Sigma_{U,Y} \subseteq \Sigma_{\sigma,M}$. This motivates the following definition.
\begin{definition}\label{def_inf1}
	The data set $(U,Y)$ is informative for interpolation at $\sigma$ if there exists a unique $M$ such that $\Sigma_{U,Y} \subseteq \Sigma_{\sigma,M}$.
\end{definition}

Note that the condition $\Sigma_{U,Y} \subseteq \Sigma_{\sigma,M}$ asks for all systems explained by the data to have the same transfer function value $M$ at $\sigma$. It is clear that, if the input data $U$ is persistently exciting of a sufficient order, then the data is sufficient to identify the system, i.e., $\Sigma_{U,Y}=\{\begin{bmatrix}
\bar{q} & -\bar{p}
\end{bmatrix}\}$, and therefore such unique $M$ exists (in case $\bar{P}(\sigma) \neq 0$). In this extended abstract, we however look for conditions for informativity for interpolation that are strictly weaker than those for identification. This motivates the following problem statement. 
\begin{problem}\label{problem2}
	Find necessary and sufficient conditions such that $(U,Y)$ is informative for interpolation at $\sigma$. Furthermore, if the data satisfies these conditions, then find $M$.
\end{problem}

\section{Data Informativity for Interpolation}\label{Section_result}
In this section, we will provide necessary and sufficient conditions under which the data is informative for interpolation at a given frequency (interpolation point).


In order to able to characterize informativity for interpolation, we first state the following theorem. 
	\begin{thm}\label{Thm_two_incl}
		Consider two sets $\Sigma_{U,Y}$ and $\Sigma_{\sigma,M}$ in (\ref{Set_UY}) and (\ref{Set_sigmaM}), respectively. The inclusion $\Sigma_{U,Y} \subseteq \Sigma_{\sigma,M}$ holds if and only if
		\begin{equation}\label{Eqn_lin2}
		\begin{bmatrix}
		H_{n}(U) & 0 \\ H_{n}(Y) & -\gamma_n(\sigma)
		\end{bmatrix} \begin{bmatrix}
		\xi \\ M
		\end{bmatrix}=\begin{bmatrix}
		\gamma_n(\sigma) \\ 0
		\end{bmatrix}
		\end{equation}
		for some $\xi \in \C^{T-n+1}$ and $M \in \C$.
\end{thm}
\textcolor{black}{
The proof of Theorem \ref{Thm_two_incl} uses linear algebra to exploit the inclusion of solutions of two distinct linear equations.}

\textcolor{black}{
Theorem \ref{Thm_two_incl} is instrumental to represent the system inclusion in Definition \ref{def_inf1}. Specifically, the inclusion $\Sigma_{U,Y} \subseteq \Sigma_{\sigma,M}$ can be represented in a linear equation whose matrices only contain the input-output data and the given interpolation point. }

The main result of this extended abstract is stated in the following theorem.
\begin{thm}\label{Thm_1}
	The data $(U,Y)$ is informative for interpolation at $\sigma$ if and only if 
	\begin{equation}\label{condition2}
	\rank\begin{bmatrix}
	H_{n}(U) & 0  & \gamma_n(\sigma) \\ H_{n}(Y)  & \gamma_n(\sigma) & 0 
	\end{bmatrix}=\rank\begin{bmatrix}
	H_{n}(U) & 0 \\ H_{n}(Y)  & \gamma_n(\sigma)
	\end{bmatrix} 
	\end{equation}
	and 
	\begin{equation}\label{condition3}
	\rank\begin{bmatrix}
	H_{n}(U) & 0 \\ H_{n}(Y) & \gamma_n(\sigma)
	\end{bmatrix} = \rank\begin{bmatrix}
	H_{n}(U)  \\ H_{n}(Y) 
	\end{bmatrix} +1 .
	\end{equation}
\end{thm}
\textcolor{black}{The proof of this theorem follows from the solvability of (\ref{Eqn_lin2}) and a condition for guaranteeing the uniqueness of $M$.  Consequently, if conditions (\ref{condition2}) and (\ref{condition3}) hold, then the unique $M$ can be retrieved by solving (\ref{Eqn_lin2}). This answers the second part of Problem \ref{problem2}. }

Theorem \ref{Thm_1} allows for efficiently verifying whether the data is informative for interpolation at $\sigma$ or not by checking the ranks of some data-based matrices. Moreover, since we give necessary and sufficient conditions, we can infer that if both conditions do not hold, then the data is not informative for the desired interpolation. Another important consequence of this theorem is that the rank-equality in (\ref{condition2}) does not require to be full rank (i.e. equal $2n+2$). Thus, it is possible to have (infinitely) many systems explained by the data, but have the same transfer function value $M$ at $\sigma$. Nonetheless, this knowledge is sufficient to obtain a reduced-order model satisfying interpolation.

\section{Illustrative Example}\label{Section_example}
\begin{figure}
	\centering
	\scalebox{0.8}{
		\begin{circuitikz}[scale=0.8]
			\node [ocirc](TW) at (-0.56,3) { }; \node [ocirc](TW) at (-0.56,0) {};
			\draw (-0.5,3)--(1,3) to [cute inductors, L=$L_1$] (3.5,3) to [cute inductors, L=$L_2$] (6,3) to [cute inductors, L=$L_3$] (8.5,3) to [cute inductors, L=$L_4$] (11,3);
			\draw (-0.5,0)-- (11,0);
			\draw (1,3) to [european resistors, R=$R_1$] (1,0);
			\draw (3.5,3) to [european resistors, R=$R_2$] (3.5,0);
			\draw (6,3) to [european resistors, R=$R_3$] (6,0);
			\draw (8.5,3) to [european resistors, R=$R_4$] (8.5,0);
			\draw (11,3) to [european resistors, R=$R_5$] (11,0);
			\node [] at (-0.8,0.35) {$+$}; \node [] at (-0.8,2.65) {$-$};
			\node[] at (-0.8,1.5) {$V_d$};
	\end{circuitikz}}
	\caption{RL circuit with four inductors and five resistors.}
	\label{Fig_RLcircuit}
\end{figure}
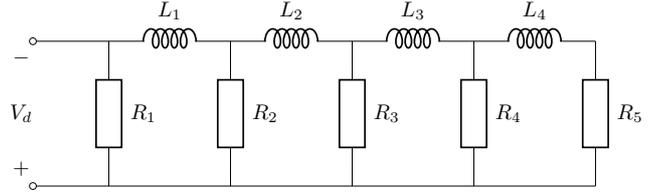

Consider the RL circuit depicted in Fig. \ref{Fig_RLcircuit}. 
The circuit is taken from \cite{jongsma2017model} with an additional inductor and resistor. We take the current through the inductors $L_1$, $L_2$, $L_3$ and $L_4$ as the state of the system, so $n=4$. The input is the voltage $V_d$. Finally, as the output, we take the current through the first inductor $L_1$. 

Let the inductances be given by $L_1=L_2=L_3=L_4=1 \ H$. For the resistors, we have $R_1=0.5\ \Omega$, $R_2=8\ \Omega$, $R_3=5\ \Omega$, $R_4=1\ \Omega$, $R_5=4\ \Omega$. As mentioned in the beginning of this abstract, we assume that the values of inductances and resistances above are unknown. Instead, we have access to the input-output data. Specifically, we collect $T=19$ samples of the input-output data using a sampling time of $0.5 \ s$, leading to the following data.
\begin{center}
	\begin{tabular}{c|c|c|c|c|c}
$t$ & $0$ & $1$ & $2$ & $3$ & $4$ \\
\hline
$U$ & $6.0000$  &  $4.8284$ &  $1.5000$ &  $-1.8284$ &  $-3.3750$  \\
\hline
$Y$ & $0$ &  $ 1.4373$  &  $2.0864$  &  $1.9180$  &  $1.1637$ \\
	\end{tabular}\\
	\begin{tabular}{c|c|c|c|c|c}
	$t$ & $5$ & $6$ & $7$ & $8$ & $9$ \\
	\hline
	$U$ & $-2.4534$  &  $0.2188$ &  $ 2.9534$ &  $4.0703$ &   $2.8675$  \\
	\hline
	$Y$ & $0.3228$ &  $-0.1050$ &  $ 0.1073$  &  $0.7958$  &  $1.5108$ \\
\end{tabular} \\

	\begin{tabular}{c|c|c|c|c|c}
	$t$ & $10$ & $11$ & $12$ & $13$ & $14$ \\
	\hline
	$U$ & $0.0215$  & $-2.8167$ &  $-3.9937$ &  $-2.8250$  &  $0.0018$  \\
	\hline
	$Y$ & $1.7858$  &  $1.4135$  &  $0.5692$  & $-0.2919$ &  $-0.7007$ \\
\end{tabular}\\

\begin{tabular}{c|c|c|c|c|c}
	$t$ & $15$ & $16$ & $17$ & $18$ & $19$ \\
	\hline
	$U$ & $2.8294$ &   $4.0005$ &    $2.8287$  &  $0.0001$  & $-2.8284$ \\
	\hline
	$Y$ & $-0.4495$  &  $0.2862$  &  $1.0504$  &  $1.3730$  &  $1.0452$
\end{tabular}
\end{center}

Suppose we aim to interpolate the transfer function values at $\sigma_1=0, \sigma_2=\frac{1}{2}, \sigma_{3,4}=\frac{1}{\sqrt{2}}\pm\frac{i}{\sqrt{2}}$, and $\sigma_5=1$. The rank conditions in Theorem \ref{Thm_1} hold for $\sigma_2, \sigma_3$, and $\sigma_4$, and thus the data $(U,Y)$ is informative for interpolation at these points. Furthermore, by solving (\ref{Eqn_lin2}), we obtain the transfer function values $M(\sigma_2)=-0.0101 - 0.2792i$, $M(\sigma_3)=-0.0101 + 0.2792i$, and $M(\sigma_4)= -0.2985$. On the contrary, the conditions in Theorem \ref{Thm_1} do not hold for $\sigma_1$ and $\sigma_5$, and the data $(U,Y)$ is not informative for interpolation at these points. In this case, we do not have a unique $M$ at each of these points.

Hence, from the informativity verifications above, we obtain the pairs $\PP=\{(\sigma_j,M(\sigma_j)), j=2,3,4\}$. Now, we use a classical interpolation method to obtain the reduced order model satisfying $\PP$. Let the reduced order model be of order $r=1$, then the use of the method in \cite{antoulas1990minimal} results in the system
 \begin{equation}\label{sys_red}
 y_{t+1}   -1.0790y_t=0.1367 u_{t+1} +  0.1045u_t.
 \end{equation}
We can readily check that the reduced-order system (\ref{sys_red}) satisfies $\PP$. 

\textcolor{black}{In this example, we consider a sinusoidal input signal. However, this data informativity concept enables to check the informativity for any given input shape. It is also worth to mention that persistence of excitation of the input is a sufficient condition for this data informativity for interpolation}.
\section{Conclusion and Future Works}\label{Section_conclusion}
Motivated by data-driven model reduction, we propose the concept of informativity for interpolation of input-output data. The main results are necessary and sufficient conditions based on the rank of the Hankel matrix of the data and given interpolation points for computing the transfer function values at these points.
Finding the reduced-order model satisfying the interpolation conditions can be regarded as a classical rational interpolation problem.

Future research directions include expanding the informativity for moment matching, i.e. the higher-order derivative of the transfer function and studying this framework to obtain the reduced-order model without computing the transfer function values and/or moments.

\bibliography{ifacconf}             
                                                   







\end{document}